\pdfoutput=1  % Instructs arXiv to run this with pdflatex rather than latex.

\documentclass[aps,prd,superscriptaddress,floatfix,nofootinbib,preprintnumbers,eqsecnum,twocolumn]{revtex4-1}

\usepackage[normalem]{ulem}
\usepackage{amsmath,amssymb,float,graphicx,placeins,multirow}
\usepackage{enumerate,slashed,cancel,comment,color,bbm,rotating,placeins,mathtools,multirow,hhline}
\usepackage{array}
\usepackage{empheq}
\usepackage{xkcdcols} % color names
\usepackage{hyperref}
\usepackage{lipsum}
\hypersetup
{
  colorlinks,
  linkcolor={blue!80!black},
  citecolor={blue!80!black},
  urlcolor={blue!80!black}
}

\newcommand{\nbox}{{\,\lower0.9pt\vbox{\hrule \hbox{\vrule height 0.2 cm
\hskip 0.2 cm \vrule height 0.2 cm}\hrule}\,}}

\newcommand{\eg}{{\em e.g.}}
\newcommand{\ie}{{\em i.e.}}

\newcommand{\beq}{\begin{equation}}
\newcommand{\eeq}{\end{equation}}
\newcommand{\lsim}{\lower.7ex\hbox{{\il{\;\stackrel{\textstyle<}{\sim}\;}}}}
\newcommand{\gsim}{\lower.7ex\hbox{{\il{\;\stackrel{\textstyle>}{\sim}\;}}}}

\newcommand{\expt}[1]{\left\langle #1 \right\rangle}
\newcommand{\il}[1]{\mbox{$#1$}}

\def\tnow{t_{\mathrm{now}}}

\def\rhobar{{\bar{\rho}}}

\def\wth{{\widetilde{\theta}}}
\def\wtk{{\widetilde{k}}}
\def\wtg{{\widetilde{\Gamma}}}

\def\wtbr{\widetilde{\overline{\rho}}}
\def\wbar{\overline{w}}
\def\barOmega{\overline{\Omega}}

\graphicspath{{./figs}}

\begin{document}

\title{Spotting Stasis in Cosmological Perturbations}

%%======================================%%
%%-------AUTHOR LIST FOR REVTEX 4-1-----%%
%%======================================%%
\def\andname{\hspace*{-0.5em}} % gets rid of "and" in author list
\author{Keith R. Dienes}
\email[Email address: ]{dienes@arizona.edu}
\affiliation{Department of Physics, University of Arizona, Tucson, AZ 85721 USA}
\affiliation{Department of Physics, University of Maryland, College Park, MD 20742 USA}
\author{Lucien Heurtier} \email[Email address: ]{lucien.heurtier@kcl.ac.uk}
\affiliation{Theoretical Particle Physics and Cosmology Group, Physics Department, 
King's College London, Strand, London WC2R 2LS, United Kingdom}
\author{Daniel Hoover}
\email[Email address: ]{hooverdj@lafayette.edu}
\affiliation{Department of Physics, Lafayette College, Easton, PA 18042 USA}
\author{Fei Huang}
\email[Email address: ]{fei.huang@weizmann.ac.il }
\affiliation{Department of Particle Physics and Astrophysics, Weizmann Institute of Science, Rehovot 7610001, Israel}
\author{\\ Anna Paulsen}
\email[Email address: ]{anna\_paulsen@brown.edu}
\affiliation{Brown Theoretical Physics Center, Brown University, Providence, RI 02912, USA}
\author{Brooks Thomas}
\email[Email address: ]{thomasbd@lafayette.edu}
\affiliation{Department of Physics, Lafayette College, Easton, PA 18042 USA}

%\preprint{\JDK{preprint numbers?}}

\begin{abstract}
As discussed in a number of recent papers, cosmological stasis is a phenomenon wherein 
the abundances of multiple cosmological energy components with different equations of 
state remain constant for an extended period despite the expansion of the universe.  
One of the most intriguing aspects of the stasis phenomenon is that it can give rise to 
cosmological epochs in which the effective equation-of-state parameter $\langle w \rangle$ 
for the universe is constant, but differs from the canonical values associated with matter, 
radiation, vacuum energy, {\it etc.}\/  Indeed, during such a stasis epoch, 
the spatial average of the energy density of the universe evolves in precisely the same 
manner as it would have evolved if the universe were instead dominated by a perfect fluid 
with an equation-of-state parameter equal to $\langle w\rangle$.
However, as we shall demonstrate, this equivalence is broken at the level of the 
{\it perturbations}\/ of the energy density.  To illustrate this point, 
we consider a stasis epoch involving matter and radiation and demonstrate that within 
this stasis background the density perturbations associated with a 
spectator matter 
component 
 with exceedingly small energy density
exhibit a power-law growth
that persists across the entire duration of the stasis epoch.  This growth can potentially lead to significant enhancements of structure at small scales.  
Such enhancements are not only interesting in their own right, but may also provide a 
way of observationally distinguishing between a stasis epoch and an epoch of 
perfect-fluid domination --- even if the universe has the same equation of state in both cases.
\end{abstract}

\maketitle
%\tableofcontents

%========================================================================
%          KEYSTROKE-SAVING MACROS, nothing complicated
%========================================================================

%%%%%%%%%%%%%%%%%%%%%%%%%%%%%%%%%%%%%%%%%%%%%%%%%%%%%%%%%%%%%%%%%%%%%%%%%%%%%%%%
%\FloatBarrier
\section{Introduction\label{sec:Introduction}}

%%%%%%%%%%%%%%%%%%%%%%%%%%%%%%%%%%%%%%%%%%%%%%%%%%%%%%%%%%%%%%%%%%%%%%%%%%%%%%%%

It has recently been shown that many commonly studied extensions of the Standard Model 
give rise to cosmologies wherein the universe is driven by a dynamical interplay 
between particle-physics processes and cosmological expansion --- an interplay which 
propels the universe toward a state in which the abundances $\Omega_i$ of multiple 
cosmological energy components with different equation-of-state parameters $w_i$ 
lose their traditional time-dependent scaling behaviors and instead remain 
constant despite cosmological expansion.  This state is known as
{\it cosmological stasis}\/~\cite{Dienes:2021woi,Dienes:2023ziv}.  Such a stasis state emerges 
naturally in a variety of different BSM contexts~\cite{Dienes:2021woi,Barrow:1991dn,
Dienes:2022zgd,Dienes:2023ziv,Dienes:2024wnu,Halverson:2024oir,Barber:2024vui,
Barber:2024izt,Huang:2025odd}; nevertheless, all such realizations of stasis exhibit 
a common underlying structure~\cite{Barber:2024izt}.  Moreover, the modification 
of the cosmological expansion history associated with a stasis epoch can have observable 
impacts --- for example, on the cosmic microwave background and on the stochastic 
gravitational-wave background~\cite{Dienes:2022zgd}.

One particularly intriguing aspect of the stasis phenomenon is that the effective 
equation-of-state parameter $\langle w\rangle \equiv \sum_i\Omega_i w_i$ for the universe 
as a whole can take a value which is constant and yet differs from the canonical values 
associated with matter ($w_i = 0$), radiation ($w_i = 1/3$), vacuum energy ($w_i = -1$), 
kination ($w_i = 1$), and so forth.  Indeed, as far as the {\it global}\/ properties of a flat 
Friedmann-Robertson-Walker (FRW) universe are concerned --- properties such as the spatial 
average $\overline{\rho}$ of the total energy density or the Hubble parameter 
$H \equiv a^{-1}(da/dt)$ --- there is an equivalence between a stasis epoch involving many 
different cosmological energy components and an epoch dominated by a single perfect fluid 
with an equation-of-state parameter $w_{\rm PF}$ equal to $\langle w\rangle$.

Despite this fact, we shall demonstrate in this paper that this equivalence is 
broken at the level of spatial inhomogeneities.  To do this, we shall consider a stasis 
cosmology.  Within this cosmology, we shall then analyze the evolution of perturbations 
of not only the energy densities that are directly involved in the stasis dynamics, 
but also the energy densities associated with any ``spectator'' components which
couple only gravitationally to the stasis sector and whose energy densities are extremely 
small compared to the critical density during that epoch.
As a concrete example, we shall consider the case in which the collective decays of
a tower of massive particle species directly into 
massless final-state particles give rise to a matter/radiation stasis, and in which a spectator 
matter component is also present.  We then demonstrate that once they enter the horizon, density perturbations for this spectator 
matter component exhibit a power-law growth that persists across the entire duration of the stasis epoch.
This behavior stands in sharp contrast to that which such perturbations exhibit during 
an epoch of perfect-fluid domination, in which such perturbations experience 
only a short transient phase of growth 
and then plateau~\cite{Redmond:2018xty}.
This power-law growth that emerges during stasis also differs from the linear growth that such 
perturbations exhibit~\cite{Erickcek:2011us} during an early matter-dominated era (EMDE).~ 
Nevertheless, in situations in which this spectator component plays the role of the dark matter, 
this enhanced growth of perturbations during stasis can potentially lead to an enhancement 
of structure at small scales.  Such an enhancement could potentially provide observational 
handles which might allow one to distinguish a stasis epoch from an epoch of 
perfect-fluid domination in the early universe --- even when the universe in each 
scenario has the same equation of state.

This paper is organized as follows.  First, in Sect.~\ref{sec:eoms}, we review
the general formalism which describes the evolution of density perturbations 
in an FRW universe.  Then, in Sect.~\ref{sec:fluid}, we review how density perturbations
associated with a spectator matter component evolve during an epoch of perfect-fluid domination.
Finally, in Sect.~\ref{sec:stasis}, we investigate how such perturbations evolve during 
a matter/radiation stasis epoch and demonstrate that they experience power-law growth 
after entering the horizon.  We conclude in Sect.~\ref{sec:conc}
with a summary of our results and a discussion of possible directions for future work.

%%%%%%%%%%%%%%%%%%%%%%%%%%%%%%%%%%%%%%%%%%%%%%%%%%%%%%%%%%%%%%%%%%%%%%%%%%%%%%%%
%\FloatBarrier
\section{Perturbation Evolution\label{sec:eoms}}

%%%%%%%%%%%%%%%%%%%%%%%%%%%%%%%%%%%%%%%%%%%%%%%%%%%%%%%%%%%%%%%%%%%%%%%%%%%%%%%%

In general, the manner in which the spacetime metric $g_{\mu\nu}$ and the energy densities 
$\rho_X$ of a set of cosmological energy components $X$ behave as functions of the spacetime 
coordinates $x^\mu$ may be ascertained from the Einstein 
equation and the equations $\nabla_\mu (T_X)^{\mu}_{~\,\nu} = (Q_X)_\nu$ which
describe the transfer of stress-energy between cosmological components, where the 
source and sink terms $(Q_X)_\mu$ are subject to the stress-energy-conservation 
condition $\sum_X \nabla_\mu (T_X)^{\mu}_{~\,\nu} = 0$, or equivalently
$\sum_X (Q_X)_\nu=0$.  We focus on the case in which the $X$ behave like perfect fluids, with
\begin{equation}
  (T_X)^{\mu}_{~\,\nu} ~=~ (\rho_X + p_X)(u_X)^\mu (u_X)_\nu 
    + p_X\delta^{\mu}_{\nu}~,    
\end{equation}
where $p_X$ is the pressure associated with $X$, 
where $(u_X)^\mu$ is its contravariant four-velocity, and where 
$\delta^{\mu}_{\nu}$ denotes the Kronecker delta. 

At the global level --- \ie, neglecting
inhomogeneities --- we assume that the metric is that of an FRW 
universe with vanishing spatial curvature.  Since we are primarily interested in how 
density perturbations evolve, we consider only scalar perturbations to $g^{\mu\nu}$.  
In the conformal Newtonian gauge, the most general form of the perturbed metric involving 
only perturbations of this sort is
\begin{equation}
  ds^2 ~=~ -a^2(1+2\Psi)d\tau^2 + 
    a^2(1+2\Phi)\delta_{ij}dx^idx^j~,
  \label{eq:PertMetric}
\end{equation}
where $x^0 = \tau$ is the conformal time, where $x^i$ with $i = \{1,2,3\}$ denoting
the three spatial coordinates, where $a$ is the scale factor, where 
$\Psi(\tau,\vec{\mathbf{x}})$ is the Newtonian potential, and where 
$\Phi(\tau,\vec{\mathbf{x}})$ is the spatial curvature.  In the absence of 
anisotropic stress, $\Psi = -\Phi$.

We decompose the energy density $\rho_X(\tau,\vec{\mathbf{x}}) = 
\rhobar_X(\tau) + \Delta \rho_X(\tau,\vec{\mathbf{x}})$ of each cosmological 
component in terms of its spatial average $\rhobar_X(\tau)$ and a position-dependent 
perturbation $\Delta \rho_X$.  The pressure $p_X(\tau,\vec{\mathbf{x}}) = 
\overline{p}_X(\tau) + \Delta p_X(\tau,\vec{\mathbf{x}})$ and
each spatial component $(u_X)^i(\tau,\vec{\mathbf{x}}) = (\overline{u}_X)^i(\tau) 
+ (v_X)^i(\tau,\vec{\mathbf{x}})$ of $(u_X)^\mu$ may be decomposed 
in an analogous manner.  Given that the universe is observed to be isotropic on 
large distance scales, we take $(\overline{u}_X)^\mu = 0$ for all $X$.  We define 
the abundance $\Omega_X$ and equation-of-state parameter $w_X$ for each $X$ via the relations 
$\Omega_X \equiv \rhobar_X/\rho_{\rm crit}$ and $\overline{p}_X = w_X\rhobar_X$, 
respectively, where $\rho_{\rm crit}\equiv 3H^2/(8\pi G)$ denotes the critical density.  
We also define the fractional overdensity (or ``density contrast'') 
$\delta_X \equiv \Delta\rho_X/\rhobar_X$ as well as 
the divergence $\theta_X \equiv \partial_i(v_X)^i$ of the three-velocity for each $X$.  

In the regime wherein the perturbations are all sufficiently small, we may 
expand the Einstein and stress-energy-transfer equations to linear order in these 
perturbations in order to obtain equations of motion for $\delta_X$, the $\wth_X$, 
and $\Phi$ --- or, equivalently, for their Fourier transforms $\delta_{kX}$, 
$\wth_{kX}$, and $\Phi_k$, where $k$ denotes the comoving wavenumber of the Fourier 
mode.  At zeroth order in this expansion, the Einstein equation yields the Friedmann
equations while the stress-energy-transfer equation for each $X$ yields an equation
of motion
\begin{eqnarray}
  \dot{\rhobar}_X &~=~& -3\,\frac{\dot{a}}{a}(1+w_X)\rhobar_X - (\overline{Q}_X)_0
  \label{eq:eomgenrhodot}
\end{eqnarray}
for $\rhobar_X$, where a dot denotes a derivative with respect to $\tau$, and
where $(\overline{Q}_X)_\mu$ denotes the expression for $(Q_X)_\mu$ 
evaluated at zeroth order in the perturbations.

At first order in the perturbations, the $\mu = i$, $\nu = j$, and $\mu=\nu=0$ components 
of the Einstein equation respectively yield the relations 
\begin{eqnarray}
  4\pi Ga^2 \sum_X \delta_{X}  \rhobar_X c_{sX}^2 &\,=\,& -\ddot{\Phi} 
    - 3\frac{\dot{a}}{a}\dot{\Phi}
    - \left(2\frac{\ddot{a}}{a} - \frac{\dot{a}^2}{a^2}\right)\Phi \nonumber \\
  4\pi Ga^2 \sum_X \delta_X\rhobar_X &\,=\,& 3\frac{\dot{a}}{a}\dot{\Phi}
    + 3\frac{\dot{a}^2}{a^2}\Phi -\nabla^2\Phi ~,
  \label{eq:G00andGijEqsPsiSub}
\end{eqnarray}
where $c_{sX} \equiv \sqrt{\Delta p_X/\Delta \rho_X}$ is the effective sound speed 
of $X$.  In general, this $c_{sX}$ differs from the adiabatic sound speed.  However, 
we note that $c_{sX} \approx 0$ for a highly non-relativistic matter component, 
whereas for radiation we have $c_{s\gamma} \approx 1/\sqrt{3}$.
Indeed, the effective sound speed for a population of particles is 
related to their underlying phase-space distribution $f(\vec{\mathbf{x}},\vec{\mathbf{p}})$
by
\begin{equation}
c_{sX}^2 ~=~ \frac{\Delta P_X}{\Delta \rho_X} ~=~ \frac{
   \int d^3 p \, \Delta f_X(\vec{\mathbf{x}},\vec{\mathbf{p}}) 
   \frac{|\vec{\mathbf{p}}|^2}{3E_p}}
   { \int d^3 p \, \Delta f_X(\vec{\mathbf{x}},\vec{\mathbf{p}}) E_p}
  \label{eq:csXRationofInts}
\end{equation}
where $E_p$ is the energy of a particle of species $X$ with momentum $\vec{\mathbf{p}}$
and where $\Delta f_X(\vec{\mathbf{x}},\vec{\mathbf{p}}) 
\equiv f_X(\vec{\mathbf{x}},\vec{\mathbf{p}}) - \bar{f}_X(\vec{\mathbf{p}})$ represents 
the position-dependent perturbation in $f_X(\vec{\mathbf{x}},\vec{\mathbf{p}})$ away 
from its spatial average $\bar{f}_X(\vec{\mathbf{p}})$.
Within the regime wherein $f_X(\vec{\mathbf{x}},\vec{\mathbf{p}})$ and therefore
$\Delta f_X(\vec{\mathbf{x}},\vec{\mathbf{p}})$ is highly non-relativistic, the above expression
reduces to $c_s^2 \approx \expt{v^2}_{\Delta f_X}/3 \approx 0$, where 
$\expt{v^2}_{\Delta f_X}$ denotes the expectation value of the speed $v$ 
of the particles which constitute the perturbation 
$\Delta f_X(\vec{\mathbf{x}},\vec{\mathbf{p}})$ at a given location 
$\vec{\mathbf{x}}$.  Likewise, if $f_X(\vec{\mathbf{x}},\vec{\mathbf{p}})$ is highly 
relativistic, then $E_p \approx |\vec{\mathbf{p}}|$ across the population of particles and 
Eq.~(\ref{eq:csXRationofInts}) reduces to $c_s^2 \approx 1/3$.

The stress-energy-transfer equation for each $X$ yields equations of
motion for $\delta_X$ and $\theta_X$ of the form
\begin{eqnarray}
  \dot{\delta}_X &~=~& - 3\dot{\Phi}(1 + w_X)
    - (1 + w_X)\theta_X
    \nonumber \\ & & ~~+\frac{1}{\overline{\rho}_X} 
    \Big[(\overline{Q}_X)_0\delta_X - (\Delta Q_X)_0\Big] \nonumber \\
  \dot{\theta}_X &~=~& 
  -\frac{\dot{a}}{a}(1 - 3w_X)\theta_X 
  - \frac{w_X\nabla^2\delta_X}{1+w_X} + \nabla^2\Phi
  \nonumber \\ & &
 ~~ + \frac{1}{\overline{\rho}_X}
  \left[\frac{\delta^{ij}\partial_j(\Delta Q_X)_i}{1+w_X}
   + (\overline{Q}_X)_0 \theta_X \right]~,~~~
  \label{eq:FinalEOMdeltaXthetaX}
\end{eqnarray}
where $(\Delta Q_{X})_\mu$ represents the contribution to $(Q_X)_\mu$ which 
arises at first order in the perturbations.

We also note that the two equations in Eq.~(\ref{eq:G00andGijEqsPsiSub}) can be 
combined to yield a single second-order differential equation for $\Phi$, the 
Fourier transform of which is 
\begin{empheq}[box=\fbox]{align}
  ~~
  & \Phi_k'' + \frac{7+3\langle w\rangle}{2a} \,\Phi_k'
     + \frac{\langle w\rangle k^2}{a^4H^2} \Phi_k \nonumber \\ 
   & ~~~~~ 
   ~=~  \frac{3}{2a^2} \sum_X \Omega_X \delta_{kX} 
     \Big(\langle w\rangle - c^2_{sX}\Big)~,~~
  \label{eq:dPrimePhiEqnArbwSimpFTDimVars}
\end{empheq}
where a prime denotes a derivative with respect to $a$ and where
$\langle w \rangle \equiv \sum_X \Omega_X w_X$.  The right side of this 
equation can be interpreted as a source for the gravitational potential.
As we shall see, this source term plays a crucial role in the evolution 
of density perturbations during stasis. 

%%%%%%%%%%%%%%%%%%%%%%%%%%%%%%%%%%%%%%%%%%%%%%%%%%%%%%%%%%%%%%%%%%%%%%%%%%%%%%%%
%\FloatBarrier
\section{The Perfect-Fluid Case\label{sec:fluid}}

%%%%%%%%%%%%%%%%%%%%%%%%%%%%%%%%%%%%%%%%%%%%%%%%%%%%%%%%%%%%%%%%%%%%%%%%%%%%%%%%

As a warm-up, let us first consider the simple case of an epoch during which the energy 
density of the universe is dominated by a single 
perfect fluid with a constant equation-of-state parameter $w_{\rm PF}$ within the range
$0<w_{\rm PF}<1/3$.
For convenience in what follows, we shall work primarily in terms of dimensionless 
variables obtained by scaling the corresponding dimensionful quantities by an appropriate 
power of the value $H^{(0)}$ at some initial time $t^{(0)}$.  In this way, we define a 
dimensionless Hubble parameter $E \equiv H/H^{(0)}$, a dimensionless comoving 
wavenumber $\wtk \equiv k/H^{(0)}$, and dimensionless velocity gradient 
$\wth_X \equiv \theta_X/H^{(0)}$.  We also adopt the convention that 
$a_0 \equiv a(t^{(0)}) = 1$ at this initial time. 

During an epoch of perfect-fluid domination, there is effectively only one term in 
the sum over $X$ in Eq.~(\ref{eq:dPrimePhiEqnArbwSimpFTDimVars}).  Likewise, we have
$\langle w \rangle = w_{\rm PF}$.  Thus, since $c_{s{\rm PF}}^2 = w_{\rm PF}$ for such
a fluid, the right side of this equation vanishes, yielding a homogeneous equation for 
$\Phi_k$ which is independent of the other perturbations.  Accounting for boundary 
conditions at early times, we then find that the solution to 
Eq.~(\ref{eq:dPrimePhiEqnArbwSimpFTDimVars}) is given by~\cite{Redmond:2018xty}
 \begin{eqnarray}
   \Phi_k &\,=\,& \Phi_{k0} \,\Gamma(\eta+1) 
       \left(\frac{a^{\frac{3w_{\rm PF}+1}{2}}\wtk w_{\rm PF}^{1/2}}
         {1+3w_{\rm PF}}\right)^{-\eta} \nonumber \\ 
    & & ~~~~~~~~ \times 
       J_{\eta} \bigg(
      \frac{2a^{\frac{3w_{\rm PF}+1}{2}}\wtk w_{\rm PF}^{1/2}}{1+3w_{\rm PF}}\bigg)~ 
   \label{eq:PhikGenSolPFDom}
\end{eqnarray}
where $\Gamma(x)$ denotes the Euler gamma function of $x$,
where $J_\alpha(x)$ denotes the Bessel function of the first kind, and where 
$\eta \equiv (5+3w_{\rm PF})/(2+6w_{\rm PF})$.

We can also investigate how a density perturbation for a spectator 
matter component $\chi$ with $w_\chi=0$ evolves in the presence of such a perfect fluid.  
The corresponding equations of motion for $\delta_{k\chi}$ and $\wth_{k\chi}$ are given by 
Eq.~(\ref{eq:FinalEOMdeltaXthetaX}) and in this case can be written in the form
\begin{eqnarray} 
  a^2E\delta_{k\chi}' &~=~& - 3a^2E\Phi_k' - \wth_{k\chi}
    \nonumber \\
  a^2E\wth_{k\chi}' &~=~& -aE\wth_{k\chi} -\wtk^2\Phi_k~.
\label{eq:justthesetwo}
\end{eqnarray}

%=============BEGIN FIGURE=================%
\begin{figure}[t!]
    \begin{center}
    \includegraphics[width=0.49\textwidth, keepaspectratio]{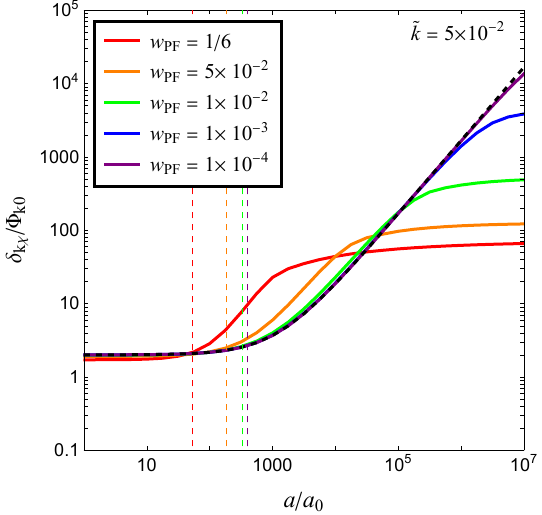}
        \caption{The density contrast $\delta_{k\chi}$ for a spectator matter
          field ($w_\chi = 0$), normalized to the value of $\Phi_{k0}$ and plotted as a 
          function of the normalized scale factor $a/a_0$ within a cosmology in 
          which a perfect fluid with a constant equation-of-state parameter $w_{\rm PF}$
          dominates the energy density of the universe.  The solid curves of different colors
          correspond to different values of $w_{\rm PF}$, while the black dashed curve 
          corresponds to the case in which $w_{\rm PF} = 0$.  For each non-zero 
          $w_{\rm PF}$, the vertical dashed line of the corresponding color indicates the 
          value of $a/a_0$ at horizon entry.
\label{fig:fig1}}
    \end{center}
\end{figure}
%===============END FIGURE=================%

These equations 
can be combined into a single second-order equation for $\delta_{k\chi}$.  Since 
$E = a^{-3(1+ \langle w\rangle)/2}$ during any epoch wherein $\langle w \rangle$ is 
constant, this equation may be expressed as
\begin{equation}
  \delta_{k\chi}'' + \frac{3}{2a}\big(1-\langle w\rangle\big)\delta_{k\chi}'  
    ~=~ S_{k\chi}~
  \label{eq:FulldeltasEqwPFConst}
\end{equation}
where we have defined
\begin{equation}
  S_{k\chi} ~\equiv~ -3\Phi_k''
    - \frac{9}{2a}\big(1-\langle w\rangle\big)\Phi_k' 
    + \wtk^2a^{3\langle w\rangle - 1}\Phi_k~.
  \label{eq:Skchi}
\end{equation}
The full solution to this inhomogeneous linear equation may be obtained via the 
method of Green's functions.  In particular, applying appropriate boundary conditions for 
$\delta_{k\chi}$ at early times and making the additional assumption that the primordial 
perturbations are purely adiabatic, we find
\begin{equation}
  \delta_{k\chi} ~=~ \frac{2\Phi_{k0}}{1 + \langle w \rangle} 
    + \int_0^a db \, G_{k\chi}(a,b) \,S_{k\chi}(b)~,
  \label{eq:GreensfnInt}
\end{equation}
where $\Phi_{k0}$ denotes the value of $\Phi_k$ at very early times, well 
before horizon entry, and where the Green's function $G_{k\chi}(a,b)$ is
given for all $0 < \langle w \rangle < 1/3$ by
\begin{equation}
  G_{ks}(a,b) ~=~ 
    \frac{2b}{1 - 3 \langle w\rangle} 
    \left[ 1 - \left(\frac{b}{a}\right)^{(1 - 3 \langle w \rangle)/2} \right]~.  
\end{equation}

In Fig.~\ref{fig:fig1}, we show how $\delta_{k\chi}$ evolves
as a function of $a$ for a mode with $\wtk = 5\times 10^{-2}$  during an epoch 
of perfect-fluid domination for a variety of different choices of 
$w_{\rm PF}$.  For all $0 < w_{\rm PF} < 1/3$, the spectator density contrast 
undergoes a transient period of growth shortly after horizon entry and then plateaus, 
as discussed in Ref.~\cite{Erickcek:2011us}.  However, we note that the transient period of 
growth which $\delta_{k\chi}$ experiences becomes longer and longer as $w_{\rm PF}$ 
decreases.  In the $w_{\rm PF} \rightarrow 0$ limit, this period of growth becomes 
infinite and we recover the behavior that we would expect for the density contrast of 
a matter component during a matter-dominated epoch --- namely, that $\delta_{k\chi}$ 
increases linearly with $a$ for all $a$.

%%%%%%%%%%%%%%%%%%%%%%%%%%%%%%%%%%%%%%%%%%%%%%%%%%%%%%%%%%%%%%%%%%%%%%%%%%%%%%%%
%\FloatBarrier
\section{Growth Through Stasis \label{sec:stasis}}

%%%%%%%%%%%%%%%%%%%%%%%%%%%%%%%%%%%%%%%%%%%%%%%%%%%%%%%%%%%%%%%%%%%%%%%%%%%%%%%%

We now proceed to our primary goal, that of calculating how perturbations in the
energy density of a spectator matter component $\chi$ evolve during a stasis epoch.  
The primary new feature in this analysis, relative to the analysis presented in 
Sect.~\ref{sec:fluid}, turns out to be the appearance of a non-zero, stasis-induced 
source term in the equation of motion for $\Phi_k$ in 
Eq.~(\ref{eq:dPrimePhiEqnArbwSimpFTDimVars}).  As we shall see, this source term 
leads to a number of surprising and important consequences.

For concreteness, we shall consider a matter/radiation stasis involving a tower of $N$ massive 
particle species $\phi_\ell$, where the index $\ell = 0,1,\ldots, N-1$ labels the species 
in order of increasing mass.  As in Refs.~\cite{Dienes:2021woi,Dienes:2023ziv}, we consider a mass 
spectrum for the $\phi_\ell$ of the form $m_\ell = m_0 + \ell^\delta \Delta m$, 
where $m_0$, $\Delta m$, and $\delta$
are free parameters.  Each $\phi_\ell$ is unstable, with a non-zero decay width $\Gamma_\ell$,
and is presumed to decay directly into a final state comprising 
exceedingly light or massless particles $\gamma$ which behave 
like radiation, with an equation-of-state parameter $w_\gamma = 1/3$.  We take
the decay widths $\Gamma_\ell$ and the initial abundances $\Omega_\ell(t_I)$ at some early 
time $t_I$ significantly smaller than the proper lifetime of the most unstable of the 
$\phi_\ell$ to scale across the tower according to power laws of the form
\begin{equation}
  \Gamma_\ell \,=\, \Gamma_0\left(\frac{m_\ell}{m_0}\right)^\gamma~,
    ~~~~
  \Omega_\ell(t_I) \,=\, \Omega_0\left(\frac{m_\ell}{m_0}\right)^\alpha ~,
\end{equation}
where $\Omega_0(t_I)$ and $\Gamma_0$ respectively represent the initial abundance and decay 
width of the lightest tower mode and where $\alpha$ and $\gamma$ are scaling exponents.  
Indeed, as discussed in Refs.~\cite{Dienes:2021woi,Dienes:2023ziv}, scaling relations 
of this sort emerge naturally in many commonly studied models of BSM physics.
Moreover, we shall assume that the cosmological population of each $\phi_\ell$ 
is already highly non-relativistic, with an equation-of-state parameter $w_\ell \approx 0$, 
by the time the particles of any of these species begin decaying appreciably.
Finally, in addition to the matter fields $\phi_\ell$ and the radiation component, 
we shall assume that our cosmology includes an additional spectator matter field $\chi$ with 
$w_\chi = 0$ and an abundance $\Omega_\chi$ which is negligible throughout the stasis 
epoch.  This field will be assumed to couple to the stasis sector --- \ie, 
to the $\phi_\ell$ fields and the relativistic particles into which they decay --- only 
through gravity.

Given this cosmological setup,
the sink term $(Q_\ell)_\mu$ for each $\phi_\ell$ 
within Eq.~(\ref{eq:eomgenrhodot}) 
takes the canonical form 
$(Q_\ell)_\mu = \Gamma_\ell(T_\ell)_{\nu\mu} (u_\ell)^\nu$ associated with particle
decay, while $(Q_\chi)_\mu = 0$.  It therefore follows from stress-energy conservation 
that $(Q_\gamma)_\mu = -\sum_{\ell = 0}^{N-1}(Q_\ell)_\mu$.  At zeroth order in the 
perturbations, Eq.~(\ref{eq:eomgenrhodot}) then yields
\begin{eqnarray}
  \rhobar'_\ell &~=~& -\frac{3}{a}\rhobar_\ell 
    - \frac{\wtg_\ell}{aE} \rhobar_\ell \nonumber \\
  \rhobar'_\gamma &~=~& -\frac{4}{a}\rhobar_\gamma 
    + \frac{1}{aE}\sum_\ell \wtg_\ell \rhobar_\ell \nonumber \\
  \rhobar'_\chi &~=~& -\frac{3}{a} \rhobar_\chi~,
\end{eqnarray}
where we have defined $\wtg_\ell \equiv \Gamma_\ell/H^{(0)}$.
As discussed in Refs.~\cite{Dienes:2021woi,Dienes:2023ziv}, an FRW universe with energy densities 
governed by such equations is dynamically driven toward a stasis in which the total matter abundance 
$\Omega_M \equiv \sum_{\ell=0}^{N-1} \Omega_\ell$ and the radiation abundance $\Omega_\gamma$ take 
fixed values $\barOmega_M$ and $\barOmega_\gamma \approx 1- \barOmega_M$ which depend on 
the values of the model parameters $\alpha$, $\gamma$, and $\delta$.  During such 
a stasis, we have $\langle w \rangle = \barOmega_\gamma/3$.

The equations of motion for the spectator perturbations $\delta_{k\chi}$ and $\wth_{k\chi}$ 
in this stasis scenario are given by Eq.~(\ref{eq:justthesetwo}), just as they are
in the perfect-fluid-dominated scenario we discussed in Sect.~\ref{sec:fluid}.~ 
Given the results in Eq.~(\ref{eq:FinalEOMdeltaXthetaX}), the corresponding equations of 
motion for the perturbations associated with each $\phi_\ell$ and with the radiation 
component $\gamma$ are given by  
\begin{eqnarray}
  a^2E\delta_{k\ell}' &~=~& - 3a^2E\Phi_k' - \wth_{k\ell}
    + a\wtg_\ell \Phi_k
   \nonumber \\
  a^2E\wth_{k\ell}' &~=~& 
    -aE\wth_{k\ell} -\wtk^2\Phi_k 
    \nonumber \\
  a^2E\delta_{k\gamma}' &~=~& - 4a^2E\Phi_k' 
    - \frac{4}{3}\wth_{k\gamma}
    \nonumber \\ & &
    +a\sum_{\ell=0}^{N-1}\wtg_\ell
    \frac{\wtbr_\ell}{\wtbr_\gamma} 
    \Big( \delta_{k\ell} -\delta_{k\gamma} - \Phi_k \Big) 
    \nonumber \\
  a^2E\wth_{k\gamma}' &~=~& 
    \frac{1}{4}\wtk^2\delta_{k\gamma} -\wtk^2\Phi_k 
    \nonumber \\ & &
    + a\sum_{\ell=0}^{N-1}\wtg_\ell
    \frac{\wtbr_\ell}{\wtbr_\gamma}
    \left(\frac{3}{4}\wth_{k\ell} - \wth_{k\gamma} \right) ~.~
  \label{eq:PertSuiteAll} 
\end{eqnarray} 
A first-order equation of motion for $\Phi_k$ may also be obtained from 
the second equation in Eq.~(\ref{eq:G00andGijEqsPsiSub}): 
\begin{eqnarray}
  a^3E^2\Phi_k' &~=~& -\frac{1}{3}\wtk^2\Phi_k
    - a^2E^2\Phi_k \nonumber \\ & &~
    +\frac{1}{2} a^2E^2 \bigg(\Omega_\gamma\delta_{k\gamma}
  + \sum_{\ell=0}^{N-1}\Omega_\ell\delta_{k\ell}\bigg)~.~~~
  \label{eq:PertSuitePhi} 
\end{eqnarray}

In order to obtain an expression for $\delta_{k\chi}$ as a function of $a$, we numerically 
integrate the coupled system of equations for $\Phi_k$, $\delta_{k\gamma}$, $\wth_{k\gamma}$, 
the $\delta_{k\ell}$, and the $\wth_{k\ell}$ in Eq.~(\ref{eq:PertSuiteAll}).  Initial conditions
for our perturbations at the beginning of the stasis epoch are established by starting our
simulation during a matter-dominated epoch at a time $t_I \ll \Gamma_{N-1}^{-1}$ during which 
the $\phi_\ell$ collectively dominate the energy density for the universe and have not 
decayed appreciably. Under the assumption that the primordial perturbations are adiabatic, 
these initial conditions are the same as those discussed in Ref.~\cite{Erickcek:2011us}
for the corresponding cosmological components in an EMDE involving a single decaying matter 
field, with all of the $\delta_{k\ell}$ given by identical expressions equal to the expression 
for the initial density contrast of that decaying field.  We then time-evolve our system toward 
stasis.  In order to establish a meaningful comparison between the results we obtain in this way
and the results for perfect-fluid domination, we choose a time at which the universe is has 
already entered stasis as the reference time $t^{(0)}$ at which we take $a = a_0$.  In what follows, 
we shall only consider modes with $\wtk \leq 1$, which enter the horizon after the stasis epoch
has begun.

\begin{figure}[t!]
    \begin{center}
    \includegraphics[width=0.47\textwidth, keepaspectratio]{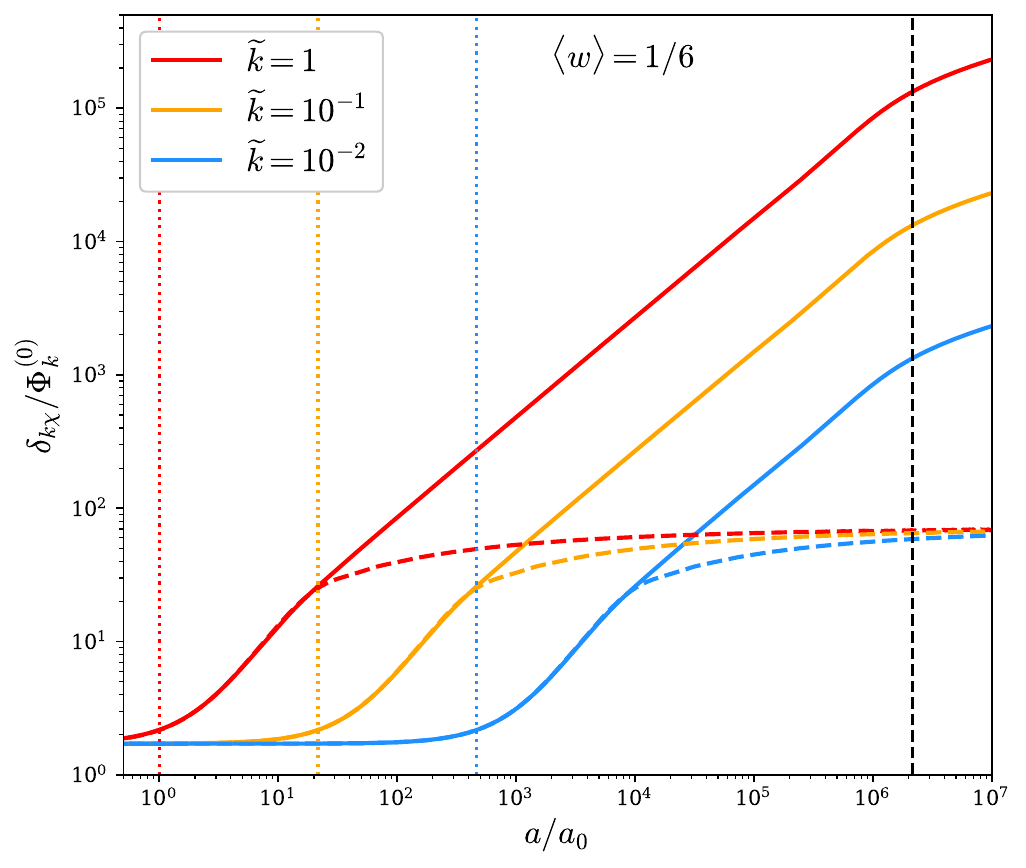}
        \caption{  
        The density contrast $\delta_{k\chi}$, plotted as a function of $a/a_0$ for a spectator 
        matter field during a matter/radiation stasis epoch with $\langle w \rangle = 1/6$
        (solid curves) and as well as during an epoch dominated by a perfect fluid 
        with $w_{\rm PF}=1/6$ (dashed curves of the corresponding color). 
        Curves are shown for three different dimensionless wavenumbers 
        $\wtk$ (different colors).  In each case, the dotted vertical line 
        indicates the time at which the corresponding mode enters the horizon, while  
        the dashed black vertical line indicates the value of $a/a_0$ at which the stasis epoch 
        ends and the subsequent epoch of radiation domination begins in the stasis cosmology.  
        We observe that after a given mode enters the horizon, its density perturbation in all 
        cases initially experiences a transient period of growth.  However, in the case of a 
        stasis epoch, $\delta_{k\chi}$ subsequently experiences sustained power-law growth 
        as long as the stasis persists.  By contrast, for the case of an epoch of perfect-fluid
        domination, $\delta_{k\chi}$ subsequently plateaus to an asymptotic value which is 
        independent of $\wtk$.  These results provide a dramatic illustration of the differences 
        between stasis and perfect-fluid domination which arise at the perturbation level.
\label{fig:fig2}}
    \end{center}

\end{figure}

In Fig.~\ref{fig:fig2} we plot $\delta_{k\chi}$ as a function of $a/a_0$ for different
values of $\wtk$ (solid curves) during a stasis epoch in which $\langle w \rangle = 1/6$.  
We observe that once each perturbation enters the horizon it begins to grow with $a$.
However, unlike during an EMDE, wherein $\delta_{k\chi}$ grows {\it linearly}\/ with 
$a$ after horizon entry~\cite{Erickcek:2011us}, we find that during stasis the spectator 
density contrast is instead described by a power law of the form 
$\delta_{k\chi}\sim a^{q(\langle w \rangle)}$, where $0 < q(\langle w \rangle) < 1$.
Indeed, for the particular choice of $\langle w \rangle$ shown in Fig.~\ref{fig:fig2},
we have $\delta_{k\chi}\sim a^{3/4}$.  

Moreover, the results shown in Fig.~\ref{fig:fig2} illustrate not only how 
$\delta_{k\chi}$ evolves during stasis, but also how it evolves at the end of
the stasis epoch, as the universe transitions from this epoch to the 
radiation-dominated epoch that follows.  Indeed, a matter/radiation stasis of 
this sort ends at a time $t \sim \Gamma_0^{-1}$, when the longest-lived of 
the $\phi_\ell$ decays into radiation and $\Omega_M$ becomes negligible.  
Moreover, as discussed in Ref.~\cite{Dienes:2021woi}, discretization effects 
which arise when only a small number of $\phi_\ell$ with lifetimes 
$\Gamma_\ell^{-1} \gtrsim t$ remain can cause the value of $\langle w\rangle$ 
to fluctuate around $\wbar$ during the last few $e$-folds of the stasis epoch.
However, the results in Fig.~\ref{fig:fig2} illustrate that $\delta_{k\chi}$ 
evolves smoothly from scaling like $\delta_{k\chi}\sim a^{q(\langle w \rangle)}$ 
to scaling like $\delta \sim \log a$ as the universe transitions from the stasis 
epoch to the subsequent radiation-dominated epoch without exhibiting similar
fluctuations.

%=============BEGIN FIGURE=================%
%===============END FIGURE=================%

The growth of $\delta_{k\chi}$ during stasis is ultimately a consequence of the source 
term on the right side Eq.~(\ref{eq:dPrimePhiEqnArbwSimpFTDimVars}).  During an epoch 
of perfect-fluid domination, only a single cosmological component has a non-negligible 
abundance, whereupon we have seen that this source term vanishes.  By contrast, during stasis, 
{\it multiple}\/ cosmological components with different equation-of-state parameters have 
non-negligible $\Omega_X$.  Moreover, the $\delta_{kX}$ for these components evolve 
differently with time.  As a result, the source term in 
Eq.~(\ref{eq:dPrimePhiEqnArbwSimpFTDimVars}) generically does {\it not}\/ vanish during 
stasis.

We may analytically determine the manner in which the power-law index $q(\langle w\rangle)$ 
depends on $\langle w \rangle$ from the fundamental 
equations of motion in Eqs.~(\ref{eq:dPrimePhiEqnArbwSimpFTDimVars}), 
(\ref{eq:PertSuiteAll}), and~(\ref{eq:PertSuitePhi}).  We begin by noting that
the $\phi_\ell$ which contribute significantly to $\Omega_M$ at each moment are those
with dimensionless decay widths $\wtg_\ell \lesssim E$, whose energy densities have not yet been 
impacted significantly by particle decay.  For all such $\phi_\ell$, the term proportional to 
$\wtg_\ell$ in the first equation in Eq.~(\ref{eq:PertSuiteAll}) is small in comparison
to the other terms and can to a reasonable approximation be neglected.  Since this is the only
term in this equation which depends meaningfully on $\ell$, it follows that the equations 
of motion for $\delta_{k\ell}$ and $\wth_{k\ell}$ for all such $\phi_\ell$ are identical.
Moreover, they are also identical in form to the corresponding equations of motion for the 
spectator perturbations $\delta_{k\chi}$ and $\wth_{k\chi}$.  This implies that
each $\delta_{k\ell}$ obeys a second-order differential equation whose form is identical to that of 
Eq.~(\ref{eq:FulldeltasEqwPFConst}).  Thus, for these $\phi_\ell$, we have
\begin{eqnarray}
  && \delta_{k\ell}'' + \frac{3}{2a}\big(1-\langle w\rangle\big)\delta_{k\ell}'  \nonumber\\
  && ~~\approx\, -3\Phi_k''  -\frac{9}{2a}\big(1-\langle w\rangle\big)\Phi_k' 
    + \wtk^2a^{3\langle w\rangle - 1}\Phi_k~.~~~~
  \label{eq:FulldeltaellsEqApprox}
\end{eqnarray}

We also note that the contributions to $\Omega_M$ from $\phi_\ell$ with 
decay widths in the opposite regime --- the regime in which $\wtg_\ell \gtrsim E$ --- are 
highly suppressed.  Thus, at any given moment, the $\phi_\ell$ which contribute meaningfully 
to the sums over $\ell$ in both Eq.~(\ref{eq:dPrimePhiEqnArbwSimpFTDimVars}) and the last 
equation in Eq.~(\ref{eq:PertSuiteAll}) are those for which $\wtg_\ell \lesssim E$,
whose $\delta_{k\ell}$ are essentially independent of $\ell$.  This implies that during
stasis, we may pull $\delta_{k\ell}$ outside of each of these sums and approximate
\begin{equation}
    \sum_{\ell=0}^{N-1} \delta_{k\ell}\Omega_\ell ~\approx~ 
    \delta_{k\ell} \barOmega_M 
      ~\approx~ \big(1-3\langle w\rangle\big)\delta_{k\ell}~.
\end{equation}

With this simplification, Eqs.~(\ref{eq:dPrimePhiEqnArbwSimpFTDimVars}), 
(\ref{eq:PertSuitePhi}), and (\ref{eq:FulldeltaellsEqApprox}) can be combined into 
a single fourth-order differential equation for $\Phi_k$.  Within the regime in 
which $\wtk \gg aE$ and in which the perturbation mode is well within the horizon, this equation 
reduces to 
\begin{eqnarray}
 0 &~\approx~& \left(\frac{\wtk}{aE}\right)^{-2} 
    \Big[a^4 \Phi''''_k + \big(10 - 3\langle w \rangle\big)a^3\Phi'''_k\Big] \nonumber \\
  &&~ + \frac{1}{3}a^2\Phi''_k 
    + \frac{9\langle w \rangle +7}{6} a\Phi'_k 
    + \frac{3\langle w\rangle^2+7\langle w\rangle}{2} \Phi_k~.\nonumber\\
    ~~~~~   
\end{eqnarray}
As $a$ increases, the higher-order terms become increasingly suppressed by the overall
prefactor $(\wtk/aE)^{-2} = \wtk^{-2}a^{-(3\langle w\rangle + 1)}$.  This fourth-order
differential equation for $\Phi_k$ then effectively reduces to a second-order equation.
For $0 < \langle w \rangle < 1/3$, the two linearly independent solutions 
$\Phi_k^{(+)}$ and $\Phi_k^{(-)}$ to this second-order equation are
given by $\Phi_k^{(\pm)} = a^{\xi_\pm(\langle w\rangle)}$, where we have defined
\begin{equation}
  \xi_\pm(w) ~\equiv~ \frac{1}{4}\Big[-9w-5\pm (9w^2-78w + 25)^{1/2}\Big]~.    
\end{equation}

At late times, when $a$ is large, the $\Phi_k^{(+)}$ solution dominates over
$\Phi_k^{(-)}$.  Thus, up to an overall normalization constant, 
we find $\Phi_k \sim \Phi_k^{(+)}$.  Indeed, inserting this expression for $\Phi_k$ into 
$S_{k\chi}$ and evaluating the integral in Eq.~(\ref{eq:GreensfnInt}), we  find 
that the spectator density contrast at late times scales with $a$ according 
to a power law of the form $\delta_{k\chi} \sim a^{q(\langle w\rangle)}$, where
\begin{equation}
  q(w) ~\equiv~ \frac{1}{4} \Big[3w-1\pm (9w^2-78w + 25)^{1/2}\Big]~. 
  \label{eq:qPowerLawIndex}
\end{equation}

%=============BEGIN FIGURE=================%
\begin{figure}[t!]
    \begin{center}
    \includegraphics[width=0.47\textwidth, keepaspectratio]{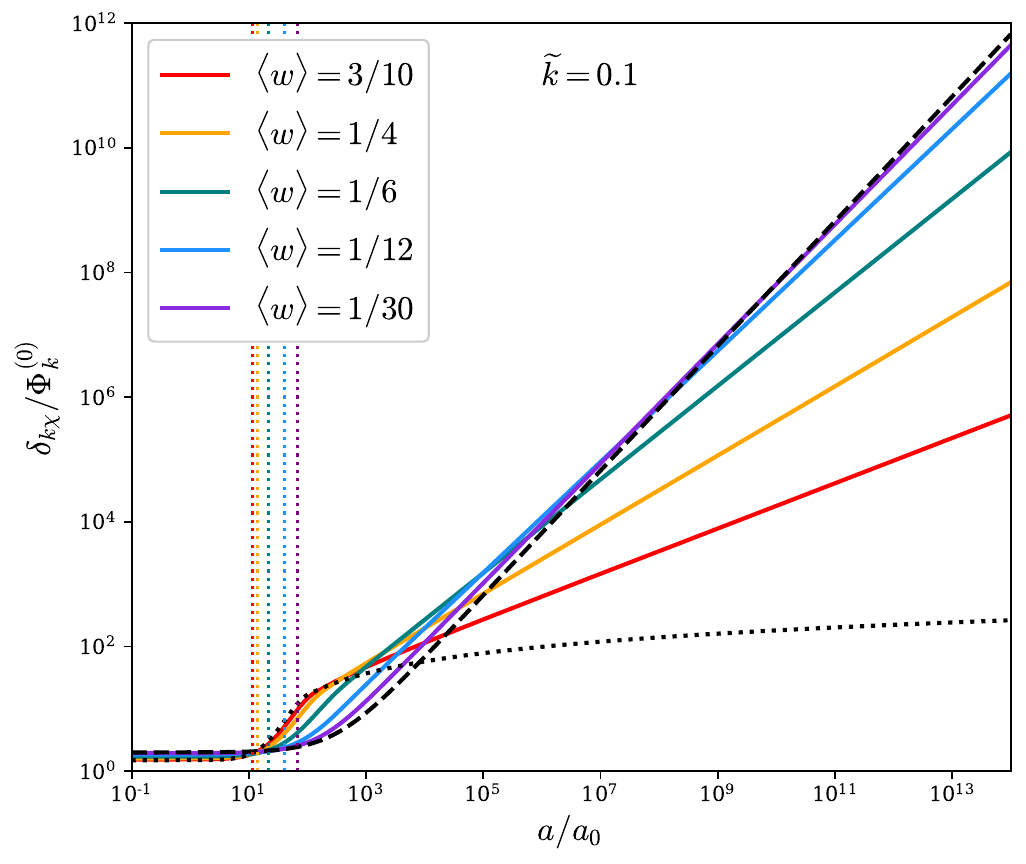}
        \caption{The density contrast $\delta_{k\chi}$ for a spectator matter field
          during stasis epochs with several different values of $\langle w \rangle$
          (solid colored curves), plotted as a function of $a/a_0$ for a perturbation 
          with $\wtk = 0.1$.  The vertical line of the corresponding color 
          indicates when the mode enters the horizon for that choice of 
          $\langle w \rangle$.  The dashed and dotted black curves indicate the 
          results for matter domination ($\langle w\rangle = 0$) and radiation 
          domination ($\langle w \rangle = 1/3$), respectively.  We observe that
          perturbations which enter the horizon during stasis exhibit power-law 
          growth at large $a$, with a power-law exponent which depends on 
          $\langle w \rangle$ in the manner indicated 
          in Eq.~(\ref{eq:qPowerLawIndex}).  We observe that for $\langle w \rangle$
          within the range $0 < \langle w \rangle < 1/3$, the behavior of 
          $\delta_{k\chi}$ interpolates between the linear and logarithmic growth
          which perturbations in matter components exhibit during matter and 
          radiation domination, respectively.
\label{fig:fig3}}
    \end{center}
\end{figure}
%===============END FIGURE=================%

In Fig.~\ref{fig:fig3}, we plot $\delta_{k\chi}$ as a function of $a/a_0$ for a mode 
with $\wtk = 0.1$ during stasis epochs with different values of $\langle w \rangle$. 
We observe that for large $a$, the spectator density contrast indeed grows with $a$ 
according to a power law with an exponent given by Eq.~(\ref{eq:qPowerLawIndex}).  
As such, for $\langle w \rangle$ within the range $0 < \langle w \rangle < 1/3$, the 
behavior of $\delta_{k\chi}$ interpolates between two extremes: the linear growth expected 
for matter perturbations during a matter-dominated epoch ($\langle w \rangle = 0$), and 
the logarithmic growth which such perturbations exhibit during a radiation-dominated epoch 
($\langle w \rangle = 1/3$).  Of course, since our initial 
conditions imply that $\delta_{k\chi} \approx 2\Phi_k^{(0)}/(1+\langle w\rangle)$ for 
modes well outside the horizon at the beginning of the stasis epoch, these curves do 
not coincide at $a/a_0 = 1$. 

It has already been noted in previous works that the matter/radiation stasis phenomenon allows 
the effective equation-of-state parameter $\langle w\rangle$ of the universe to assume 
non-traditional values within the range $0<\langle w\rangle<1/3$, with the energy density of 
the universe evolving accordingly.  However, we now find that a similar result also holds for 
the manner in which the {\it perturbations}\/ to the energy density evolve.
Thus, matter/radiation stasis represents a middle ground between matter- and radiation-domination, 
not only for the evolution of the universe as a whole but also for the evolution of its density 
perturbations.  This novel scaling behavior of $\delta_{k\chi}$ with $a$ can potentially have 
an impact on small-scale structure, with features that do not arise in either of these standard 
epochs.  However, as we have also seen, the growth of these density perturbations is wholly unlike 
that which would emerge for a universe whose energy content is dominated by the energy associated 
with a single perfect fluid, even if we could somehow arrange $w_{\rm PF}=\langle w\rangle$.  
Indeed, during a stasis epoch, the power-law growth persists throughout the duration of the stasis.  
By contrast, during an epoch dominated by a perfect fluid, the initial growth is only transient, 
with the density perturbations $\delta_{k\chi}$ quickly reaching a plateau which is independent of 
the wavenumber $\wtk$.

%%%%%%%%%%%%%%%%%%%%%%%%%%%%%%%%%%%%%%%%%%%%%%%%%%%%%%%%%%%%%%%%%%%%%%%%%%%%%%%%
%\FloatBarrier
\section{ Effect on the Matter Power Spectrum\label{sec:powerspec}}

%%%%%%%%%%%%%%%%%%%%%%%%%%%%%%%%%%%%%%%%%%%%%%%%%%%%%%%%%%%%%%%%%%%%%%%%%%%%%%%%

Given that $\delta_{k\chi}$ evolves differently during stasis than it does in the 
standard cosmology, it is interesting to consider the effect that this has on the 
linear matter power spectrum $P(k,t)$ in the case in which our spectator component 
$\chi$ plays the role of the dark matter.  In general, $P(k,t)$
is defined as the Fourier transform of the two-point correlation function 
$\xi(\vec{\mathbf{y}},t)=\langle\delta_M(\vec{\mathbf{x}},t)\delta_M(\vec{\mathbf{x}}
+\vec{\mathbf{y}},t)\rangle$, where $\delta_{M}(\vec{\mathbf{x}},t) \equiv 
[\rho_M(\vec{\mathbf{x}},t)-\overline{\rho}_M(t)]/\overline{\rho}_M(t)$ 
denotes the total perturbation in the density of matter, including contributions from 
all non-relativistic species.  This implies that up to an overall large-volume factor,
\begin{equation}
  P(k,t) ~=~ |\delta_{Mk}(t)|^2~,
  \label{eq:DefofPofk}
\end{equation}
where $\delta_{Mk}(t)$ is the Fourier transform of $\delta_{M}(\vec{\mathbf{x}},t)$.
We may also define a corresponding dimensionless power spectrum according to the usual 
convention
\begin{equation}
  \Delta^2(k,t) ~\equiv~ \frac{k^3}{2\pi^2} P(k,t)~.
  \label{eq:DimlessPowerSpectrum}
\end{equation}

Under the assumption that the spectator $\phi_s$ plays the role of the dark matter, we
may establish a relationship between $\Phi_k$ and $\delta_{k\chi}$ at late times through 
the second equation in Eq.~(\ref{eq:G00andGijEqsPsiSub}), the Fourier
transform of which may be expressed as
\begin{equation}
  \left[k^2 + 3(aH)^2\left(1+\frac{a\Phi_k'}{\Phi_k}\right)\right]\Phi_k ~=~
  4\pi G a^2 \sum_X \overline{\rho}_X \delta_{Xk}~.
  \label{eq:PrePoisson}
\end{equation}
We first consider how this equation simplifies at times 
$t_{\rm MRE} < t < t_{\rm DE}$, where $t_{\rm MRE}$ denotes the time at 
matter-radiation equality and $t_{\rm DE}$ is the time at which the universe 
transitions from matter domination to dark-energy domination.  Since the 
universe is matter-dominated while $t$ lies within this range, we can neglect 
the abundances of all species excepting those which behave like matter.  
Moreover, since we are also assuming that our perturbations are initially 
adiabatic, and therefore that the $\delta_{Xk}$ for baryonic and dark matter, 
which both have $w_X \approx 0$, are approximately equal, we have 
\begin{equation}
  \sum_X \overline{\rho}_X \delta_{Xk} ~\approx~ \overline{\rho}_M \delta_{Mk}~, 
\end{equation}
where $\overline{\rho}_M$ is the spatial average of the total matter energy 
density.

We are primarily interested in perturbation modes $\delta_{Mk}$ which have 
already entered the horizon well before the range of $t$ we are 
considering --- \ie, modes for which $k \gg aH$ 
throughout this range of $t$.  For such modes, the second term in square 
brackets in Eq.~(\ref{eq:PrePoisson}) is negligible and we obtain 
\begin{equation}
  k^2\Phi_k ~=~ 4\pi G a^2 \overline{\rho}_M \delta_{Mk}~,
  \label{eq:Poisson}
\end{equation}
which is simply the Fourier transform of the Poisson equation.   
Since the energy density of matter at such times can be written as 
\begin{equation}
  \overline{\rho}_M ~=~ 
    \frac{3H_{\rm now}^2}{8\pi G}\Omega_{M,{\rm now}} 
    \left(\frac{a}{a_{\rm now}}\right)^{-3}~, 
\end{equation}
Eq.~(\ref{eq:Poisson}) implies that
\begin{equation}
  \delta_{Mk} ~=~ \frac{2a k^2 \Phi_k}{3(aH_{\rm now})^2\Omega_{M,{\rm now}}}~.  
  \label{eq:deltakfromPoisson}
\end{equation}
Thus, in order to determine the matter power spectrum $P(k,t)$ at times within this 
range, we simply need to know the corresponding gravitational potential $\Phi_k$.

The manner in which the value of $\Phi_k$ depends on $k$ and $t$ --- or, equivalently,
on $k$ and $a$ --- for times in this range is typically parametrized as
\begin{equation}
    \Phi_k ~=~ \frac{9}{10}\Phi_k^{(p)} T(k) \frac{\mathcal{D}(a)}{a}~,
  \label{eq:PhikDecomp}
\end{equation}
where $\Phi_k^{(p)}$ is the ``primordial'' value that $\Phi_k$ would have had in
the standard cosmology immediately after the reheating epoch following inflation 
had concluded and the universe had become radiation-dominated.  The transfer function
$T(k)$ parametrizes whatever additional $k$-dependence $\Phi_k$ on $k$ develops
beyond whatever dependence was already present in $\Phi_k^{(p)}$,
while the growth $\mathcal{D}(a)$ parametrizes the dependence of $\Phi_k$ on $a$.  

The numerical prefactor in Eq.~(\ref{eq:PhikDecomp}) accounts for the fact that 
the value of $\Phi_k$ for modes with $k \lesssim aH$ experiences a shift in its 
otherwise constant value during transitions between cosmological epochs with 
different values of $\langle w \rangle$.  This shift can be determined from the 
relationship between $\Phi_k$ and the corresponding comoving curvature perturbation 
$\mathcal{R}_k$, which is conserved while $k \lesssim aH$.  In particular, prior to 
horizon entry, these quantities are related by
\begin{equation}
  \mathcal{R}_k ~=~ \frac{5+3\langle w\rangle}{3+3\langle w\rangle }\Phi_k~.
  \label{eq:CurvPertConserve}
\end{equation}
Since the primordial $\Phi_k^{(p)}$ is defined during a radiation-dominated epoch,
it then follows that modes which enter the horizon during a matter-dominated epoch 
have $\Phi_k = (9/10)\Phi_k^{(p)}$ 
Thus, our choice of numerical prefactor in
Eq.~(\ref{eq:PhikDecomp}) ensures that modes which enter the horizon at times 
$t_{\rm MRE} < t < t_{\rm DE}$ have $T(k) = 1$.

Substituting the result in Eq.~(\ref{eq:PhikDecomp}) into 
Eq.~(\ref{eq:deltakfromPoisson}), we have
\begin{equation}
  \delta_{Mk} ~=~ \frac{3 k^2 \Phi_k^{(p)}}
    {5(aH_{\rm now})^2\Omega_{M,{\rm now}}} 
    T(k)\mathcal{D}(a)~.
  \label{eq:TransFuncParam}
\end{equation}
This is the usual parametrization for $\delta_{Mk}$, and it is the one used in 
Ref.~\cite{Erickcek:2011us}.  Cosmic-microwave-background (CMB)
measurements~\cite{Planck:2018nkj} indicate that $P(k,t_{\rm now}) \sim k$ for 
modes which enter the horizon at times $t > t_{\rm MRE}$.
Thus, for an approximately scale-invariant dimensionless power spectrum of 
primordial curvature perturbations after inflation which obeys a scaling relation of 
the form
\begin{equation}
  \Delta^2_\mathcal{R}(k) ~\equiv~ \frac{k^3}{2\pi^2} \langle|\mathcal{R}_k|^2\rangle 
    ~=~ A_\ast \left(\frac{k}{k_\ast}\right)^{n_s - 1}    
\end{equation}
with a spectral index $n_s$ very near unity, where $k_\ast$ is a suitably chosen pivot scale
and where $A_\ast$ is a dimensionless constant, 
Eq.~(\ref{eq:CurvPertConserve}) implies 
that $\Phi_k^{(p)} \sim k^{-3/2}$.  For the choice $k_\ast = 0.05$~Mpc$^{-1}$, CMB 
measurements yield $n_s \approx 0.965$ and 
$A_\ast \approx 2.099\times 10^{-9}$~\cite{Planck:2018vyg}.  Thus, Eqs.~(\ref{eq:DefofPofk}) 
and~(\ref{eq:TransFuncParam}) together imply that 
\begin{equation}
  P(k,t_{\rm now}) ~\sim~ k^4 |\Phi_k^{(p)}|^2 T^2(k) 
    ~\sim~ k T^2(k) 
\end{equation}
for modes which enter the horizon at times $t > t_{\rm MRE}$.  Consistency with CMB data
therefore implies that $T(k)$ is independent of $k$ for such modes, as desired.

We emphasize, however, that $\Phi_k^{(p)}$ is not
the same as the initial value $\Phi_{k}^{(0)}$ for the corresponding perturbation mode
in our stasis cosmology.  Rather, if we choose $t^{(0)}$ to be a time at which the universe 
is already in stasis, Eq.~(\ref{eq:CurvPertConserve}) implies that
\begin{equation}
    \Phi_k^{(0)} ~=~ \frac{9(1+\wbar)}{2(5+3\wbar)} \Phi_k^{(p)}~.
\end{equation}
By contrast, if we choose $t^{(0)}$ to be a time at which the stasis epoch has not yet begun 
and the universe is dominated by the $\phi_\ell$, which behave like massive matter, then 
$\Phi_k^{(0)} = (9/10)\Phi_k^{(p)}$.

Within the context of the standard cosmology, perturbations in the energy density 
of cold dark matter --- here identified with $\delta_{k\chi}$ --- are to a good 
approximation described by a function of the form~\cite{Hu:1995en}
\begin{equation}
  \delta_{k\chi}^{{(\rm SC})} ~=~ A_{\rm SC} \Phi^{(p)}_k 
    \log\left(B_{\rm SC}\frac{a}{a_k}\right)      
  \label{eq:deltaCDMHuParamRD}
\end{equation}
at times $t \ll t_{\rm MRE}$, where $A_{\rm SC} \approx 9.11$ and 
$B_{\rm SC} \approx 0.594$ are constants and where $a_k$ denotes the value of 
$a$ at which the condition $k = aH$ is satisfied and the mode enters the horizon.  
At times $t_{\rm MRE} < t \ll t_{\rm BD}$ after the time of matter-radiation equality 
but before the time $t_{\rm BD}$ of baryon decoupling, these perturbations are 
described by a function which can be expressed as
\begin{equation}
  \delta_{k\chi}^{({\rm SC})} ~=~ \frac{3}{2}A_{\rm SC} \Phi^{(p)}_k f_1 
    \log\left(\!B_{\rm SC}\left(\frac{4}{e^3}\right)^{f_2/f_1}
    \!\frac{a_{\rm MRE}}{a_{k}}\!\right) \mathcal{D}(a)~, 
  \label{eq:deltaCDMHuParamMD}
\end{equation}
where $e$ denotes the natural log base and where the constants
\begin{eqnarray}
  f_1 &~\equiv~& 1 - 0.568 f_b + 0.094 f_b^2 \nonumber \\   
  f_2 &~\equiv~& 1 - 1.156 f_b + 0.149 f_b^2 - 0.074 f_b^3
\end{eqnarray} 
are defined in terms of the late-time baryon fraction 
$f_b \equiv \Omega_b/(\Omega_b + \Omega_\chi)$.  For 
$t_{\rm MRE} \ll t < t_{\rm BD}$, the growth function $\mathcal{D}(a)$ is given by
\begin{eqnarray}
  \mathcal{D}(a) &~=~& \left(\frac{a_{\rm MRE}}{a+a_{\rm MRE}}\right)^{\alpha} \nonumber \\
    && ~~~\times\, {_2}F_1\left(\alpha, \alpha + \frac{1}{2}; 2\alpha + \frac{1}{2}; 
    \frac{a_{\rm MRE}}{a+a_{\rm MRE}}\right)~,\nonumber \\ 
\end{eqnarray}
where ${_2}F_1(a,b;c;z)$ denotes Gauss's hypergeometric function and where
\begin{equation}
  \alpha ~\equiv~ -\frac{1}{4}\left[\sqrt{1+24(1-f_b)}-1\right]~.
\end{equation}

Comparing Eqs.~(\ref{eq:deltaCDMHuParamMD}) and~(\ref{eq:TransFuncParam}), one finds
that in the standard cosmology, the transfer function takes the form
\begin{eqnarray}
  T_{\rm SC}(k) &~=~& \frac{5A_{\rm SC} f_1}{2k^2} 
    (aH)_{\rm now}^2\Omega_{M,{\rm now}} 
    \nonumber \\
    & & ~~~\times
    \log\left(B_{\rm SC}\frac{a_{\rm MRE}}{a_{k}}\left(\frac{4}{e^3}\right)^{f_2/f_1}
    \right)~.~~~~
\end{eqnarray}
Within the context of the standard cosmology, wherein the universe is radiation dominated
between the time $t_{\rm RH}$ at the end of reheating and the time of matter-radiation 
equality, the Friedmann equation implies that
\begin{equation}
  \frac{a_{\rm MRE}}{a_k} ~=~ \frac{\sqrt{2}k}{k_{\rm MRE}}
\end{equation}
for modes which enter the horizon at times $t \ll t_{\rm MRE}$, where 
$k_{\rm MRE} \equiv a_{\rm MRE} H_{\rm MRE}$ denotes the wavenumber of the mode 
which enters the horizon at $t = t_{\rm MRE}$.  Thus, our final expression for 
$T(k)$ for short-wavelength modes in the standard cosmology is 
\begin{eqnarray}
  T_{\rm SC}(k) &~=~& \frac{5A_{\rm SC} f_1}{2k^2}
    (aH_{\rm now})^2\Omega_{M,{\rm now}} \nonumber \\
    & & ~~\times
    \log\left(\frac{\sqrt{2} B_{\rm SC} k}{k_{\rm MRE}}
    \left(\frac{4}{e^3}\right)^{f_2/f_1} \right)~.~~~~
  \label{eq:TSCFinal}
\end{eqnarray}

By contrast, within the context of non-standard cosmological scenarios, 
$\delta_{k\chi}$ may scale differently with $a$ at times $t < t_\ast$, where $t_\ast$ 
denotes the time after which the expansion history is equivalent to that of the 
standard cosmology.  Nevertheless, at times $t_\ast < t$, they scale with $a$ in the 
same manner as they do in the standard cosmology.  For example, they scale as 
$\delta_{k\chi} \propto \log(a)$ at times $t_\ast < t < t_{\rm MRE}$.  Thus, without 
loss of generality, $\delta_{k\chi}$ may be parametrized in such cosmologies 
by a function of the same form as Eq.~(\ref{eq:deltaCDMHuParamRD}) prior to 
matter-radiation equality, but with the constants $A_{\rm SC}$ and $B_{\rm SC}$ replaced 
by general functions $A(k)$ and $B(k)$ of $k$:
\begin{equation}
  \delta_{k\chi} ~=~ A(k) \Phi^{(p)}_k \log\left(B(k)\frac{a}{a_k}\right)~.      
  \label{eq:deltaStasisHuParamRD}
\end{equation}
During the later period wherein $t_{\rm MRE} \ll t < t_{\rm BD}$, the evolution
of $\delta_{k\chi}$ will accordingly be described by a function of the same general 
form as Eq.~(\ref{eq:deltaCDMHuParamMD}), but with $A_{\rm SC}$ and $B_{\rm SC}$ 
likewise replaced by $A(k)$ and $B(k)$.  The transfer function in 
our stasis cosmology therefore takes the form
\begin{eqnarray}
  T(k) &~=~& \frac{5A(k) f_1}{2k^2} (aH)_{\rm now}^2\Omega_{M,{\rm now}}
    \nonumber \\
    & & ~~\times
    \log\left(B(k)\frac{a_{\rm MRE}}{a_k}
    \left(\frac{4}{e^3}\right)^{f_2/f_1} \right)~.~~~~
  \label{eq:TFinal}
\end{eqnarray}

The forms of $A(k)$ and $B(k)$ can be determined as follows.  
Modes which enter the horizon at times $t_s < t < t_{\rm MRE}$, where $t_s$ denotes
the time at which stasis effectively ends, evolve exactly as
they evolve within the standard cosmology.  Thus, for such modes, $A(k)$ and $B(k)$ 
are independent of $k$ and given by the constants $A$ and $B$.  By contrast,
we have seen that modes which enter the horizon during stasis grow with $a$ like 
a power law of the form $\delta_{k\chi} \sim a^{q(\langle w\rangle)}$, with 
$q(\wbar)$ given by Eq.~(\ref{eq:qPowerLawIndex}), until stasis ends.  The 
normalization for these $\delta_{k\chi}$ follows from the fact that
$\delta_{k\chi} \approx 2\Phi_k^{(0)}/(1+\wbar) = 
9\Phi_k^{(p)}/(5+3\wbar)$ for each such mode after the stasis epoch has begun but
prior to horizon entry.  Thus, we find that the growth of these modes during the 
radiation-dominated epoch can be modeled as 
\begin{equation}
  \delta_{k\chi} ~\approx~ \frac{9\Phi_k^{(p)}}{5+3\wbar}
    \left(\frac{a_s}{a_k}\right)^{q(\wbar)} 
    \left[1+\log\left(\frac{a}{a_s}\right)\right]~,
\end{equation}
where $a_s \equiv a(t_s)$.  Thus, noting that
\begin{equation}
  \frac{a_s}{a_k} ~=~ \left(\frac{k}{k_s}\right)^{2/(1+3\wbar)}~,
\end{equation}
where $k_s$ denotes the wavenumber of the mode which enters the horizon
at $t = t_s$, we find that 
\begin{eqnarray}
  A(k) &~=~& \frac{9}{5+3\wbar}\left(\frac{k}{k_s}\right)^{2q(\wbar)/(1+3\wbar)} 
    \nonumber \\
  B(k) &~=~& e \left(\frac{k}{k_s}\right)^{-2/(1+3\wbar)}~.
  \label{eq:AkandBkSpecific}
\end{eqnarray}
Moreover, we note that the ratio $a_{\rm MRE}/a_k$ which appears in 
Eq.~(\ref{eq:TFinal}) can be expressed as 
\begin{equation}
  \frac{a_{\rm MRE}}{a_k} ~=~ \frac{\sqrt{2} k_s}{k_{\rm MRE}} 
    \left(\frac{k}{k_s}\right)^{2/(1+3\wbar)}
  \label{eq:aMREOverakStasis}
\end{equation}

One additional effect which can modify $T(k)$ at large $k$ is the free streaming of
dark-matter particles out of overdense regions.  This effect depends on the
primordial velocity distribution for the cosmological population of $\chi$ particles 
that plays the role of the dark matter in this scenario.  Since the production of 
$\chi$ particles is unrelated to the stasis epoch, the primordial velocity 
distribution of the $\chi$ particles can take a variety of forms.  In what follows, 
we shall simply assume that this velocity distribution is unimodal and reasonably narrow,
such that the effect that the free-streaming of $\chi$ particles has on structure formation
is well characterized by the average velocity $\langle v_\chi\rangle$ of these particles.  
In particular, for such a velocity distribution, the free-streaming length at a given value of
the scale factor $a$ is
\begin{eqnarray}
  \lambda_{\rm fs}(a) & ~=~ &
    \frac{2a_s\langle v_{\chi s}\rangle}{(aH)_{\rm now} 
    \Omega_{\gamma,{\rm now}}^{1/2}} \left(\frac{a_s}{a_{\rm now}}\right)
    \nonumber \\
      && \times
      \left[{\rm arcsinh}\left(\!\sqrt{\frac{a_{\rm MRE}}{a_s}}\right)
      - {\rm arcsinh}\left(\!\sqrt{\frac{a_{\rm MRE}}{a}}\right) \right]\,, \nonumber \\
\end{eqnarray}
where $\langle v_{\chi s}\rangle$ denotes the value of $\langle v_\chi \rangle$ at $t = t_s$
and where $\Omega_{\gamma,{\rm now}}$ denotes the present-day abundance of radiation.  
At late times, when $a \gg a_{\rm MRE}$, this expression becomes approximately independent 
of $a$.  We model the effect of free streaming on $P(k,t)$ at late times by incorporating an 
additional exponential suppression factor into our expression for $T(k)$ in 
Eq.~(\ref{eq:TFinal}) such that it becomes
\begin{eqnarray}
  T(k) &~=~& \frac{5A(k) f_1}{2k^2} a_{\rm now}^2H_{\rm now}^2\Omega_{M,{\rm now}}
  \,  e^{-\frac{k^2}{
  2k_{\rm fs}^2(a_{\rm now})}} \nonumber \\
    & & ~~~~\times
    \log\left(B(k)\frac{a_{\rm MRE}}{a_k}
    \left(\frac{4}{e^3}\right)^{f_2/f_1} \right) ~,~~~~
  \label{eq:TFinalwFS}
\end{eqnarray}
where we have defined $k_{\rm fs}(a) \equiv \lambda_{\rm fs}^{-1}(a)$.

For any primordial spectrum of perturbations $\Phi_k^{(p)}$ to the gravitational 
potential, the matter power spectrum in our stasis cosmology can be parametrized as
\begin{equation}
  P(k,t_{\rm now}) ~=~ R^2(k) \,P_{\rm SC}(k,t_{\rm now})~,
\end{equation}
where $P_{\rm SC}(k,t_{\rm now})$ denotes the present-day matter power 
spectrum in the standard cosmology and where we have defined the ratio
\begin{equation}
  R(k) ~\equiv~ \frac{T(k)}{T_{\rm SC}(k)} 
    ~=~ \frac{\delta_{sk}}{\delta_{sk,{\rm SC}}}~.
\end{equation}
A similar parametrization was adopted in Ref.~\cite{Erickcek:2011us}
in order to characterize the matter power spectrum which emerges in
cosmologies involving an EMDE.  We also note that the ratio $R(k)$ accords 
with the alternative definition of the transfer function adopted in, \eg, 
Ref.~\cite{Dienes:2020bmn}.  Explicitly, Eqs.~(\ref{eq:TSCFinal}) 
and~(\ref{eq:TFinal}) imply that
\begin{equation}
    R(k) ~=~ \frac{A(k)
      \log\left(B(k)\frac{a_{\rm MRE}}{a_k}
        \left(\frac{4}{e^3}\right)^{f_2/f_1}\right)e^{-\frac{k^2}{
  2k_{\rm fs}^2(a_{\rm now})}}}
      {A_{\rm SC}\log\left(B_{\rm SC}\frac{\sqrt{2}k}{k_{\rm MRE}}
        \left(\frac{4}{e^3}\right)^{f_2/f_1}\right)}~.
\end{equation}
We emphasize that in this expression, $a_k$ denotes the scale factor at horizon entry
{\it in the stasis cosmology}\/, which in general differs from the corresponding result
in the standard cosmology.  We also note that the analytic results in 
Eqs.~(\ref{eq:AkandBkSpecific}) and~(\ref{eq:aMREOverakStasis}) together imply that
\begin{equation}
  B(k)\frac{a_{\rm MRE}}{a_k} ~=~ \frac{\sqrt{2}k_s e}{k_{\rm MRE}}~.    
\end{equation}

\begin{figure}[t!]
\begin{center}
  \includegraphics[width=0.47\textwidth, keepaspectratio]{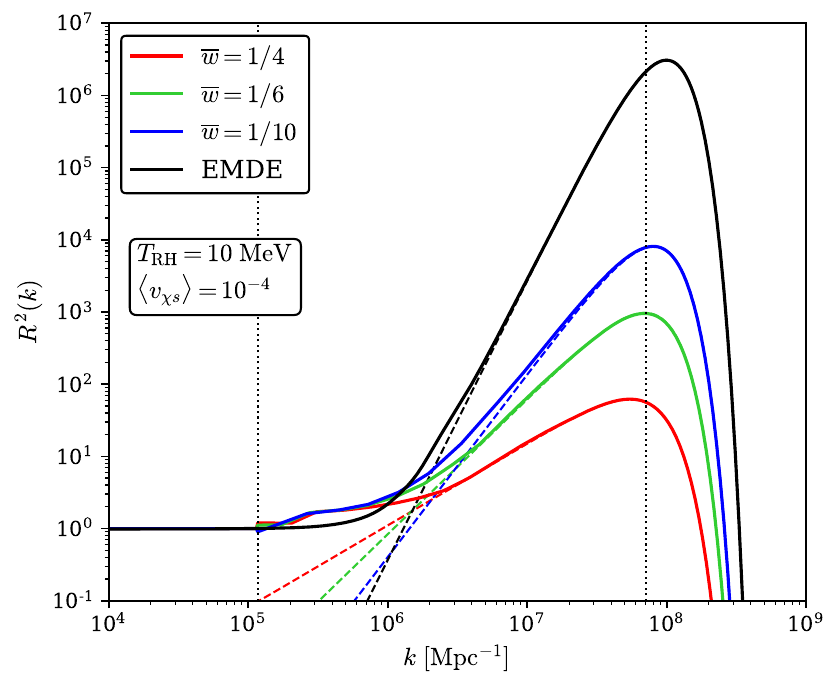}
  \caption{The factor $R^2(k)$ by which the present-day linear matter power spectrum 
    $P(k,t_{\rm now})$ is enhanced in our stasis scenario relative to the standard 
    cosmology, shown as a function of the physical wavenumber $k$ for several different 
    values of $\overline{w}$.  All curves shown correspond to the parameter choices 
    $\delta = 1$, $\gamma=7$, and $\Delta m=m_0$, with $\alpha$ determined by the value of 
    $\wbar$.  The solid curves represent our full numerical results for $A(k)$ and 
    $B(k)$.  The dashed curves represent the results obtained from the analytic expressions 
    in Eq.~(\ref{eq:AkandBkSpecific}).  The solid and dashed black curves represent the 
    corresponding enhancement factors which result from a period of early matter-domination. 
    For all curves shown, we have taken $\langle v_{\chi s}\rangle = 10^{-4}$ and 
    $T(t_s) = 10~{\rm MeV}$.  We observe that a period of cosmological stasis can lead to a
    considerable enhancement in the matter power spectrum at small $k$ and that the 
    degree of enhancement becomes larger as $\overline{w}$ decreases toward zero.
\label{fig:Rsq}}
\end{center}
\end{figure}

In Fig.~\ref{fig:Rsq}, we plot the enhancement factor $R^2(k)$ (differently colored curves) 
for the matter power spectrum which results from a period of matter/radiation stasis as a 
function of $k$ for different values of $\wbar$.  All curves shown correspond 
to the parameter choices $\delta = 1$, $\gamma=7$, and $\Delta m=m_0$, with $\alpha$ determined 
by the value of $\wbar$ in each case.  By varying $\alpha$ while holding $\delta$, $\gamma$, 
and $\Delta m$ fixed, we ensure that the mass spectrum of the tower is the same in all cases.  
The solid curves represent our full numerical results for $A(k)$ and 
$B(k)$.  The dashed curves represent the results obtained from the analytic expressions in
Eq.~(\ref{eq:AkandBkSpecific}).  The solid and dashed black curves represent the corresponding
enhancement factors which result from a period of early matter-domination.  For both stasis and 
EMDE cosmologies, we take the temperature of the SM radiation bath at $t_s$ to be 
$T(t_s) = 10$~MeV and the average velocity of dark-matter particles at this same time 
to be $\langle v_{\chi s}\rangle = 10^{-4}$.  We define $t_s$ in both the stasis and EMDE
cosmologies to be the time at which $\Omega_M$ falls below $\Omega_M=0.001$ and the universe  
may be considered radiation-dominated.  We note that this convention for the time at which 
the expansion history of the universe begins to coincide with that of the standard 
cosmology differs from the convention adopted in Ref.~\cite{Erickcek:2011us} in the 
context of a modified cosmology involving an EMDE, wherein this time was taken to be the 
canonical reheating time $t_{\rm RH}$ --- \ie, the time at which $\Omega_M\simeq 0.5$.  
The values of $k$ for the modes which cross the horizon at $t_s$ and $t_{\rm RH}$ are 
related by $k_{\rm RH}\simeq 2.85 k_{\rm s}$.

We see from Fig.~\ref{fig:Rsq} that $P(k,\tnow)$ is significantly enhanced for modes 
with wavenumbers which lie within a finite range of values --- demarcated here by the 
two dotted vertical lines.  Modes with $k$ below this range enter the horizon only after 
the end of the stasis epoch and evolve exactly as they would in the standard cosmology.  
By contrast, modes with $k$ above this range are washed out by free-streaming effects.  
Unlike in the case of an EMDE, the behavior of the curves at the lower end of this range
is sensitive to the edge effects which result from the discrete spacing between the 
masses of the tower states.  We observe that the extent to which the matter power spectrum 
is enhanced within this range of $k$ can be quite significant and that this enhancement 
increases as $\overline{w}$ approaches zero.

%=============BEGIN FIGURE=================%
%===============END FIGURE=================%

The primary observational implication of these results is that an epoch of 
matter/radiation stasis prior to the BBN epoch can lead to the formation of 
dark-matter substructures similar to those which can arise from an epoch 
of early matter domination~\cite{Erickcek:2011us,Erickcek:2015jza,Delos:2023vfv}.  
The smallest dark-matter substructures to which experiments are currently sensitive 
are those with masses $M \sim 10^{-11}M_\odot$ (for reviews, see, \eg, 
Ref.~\cite{Bechtol:2022koa}).  Constraints on both compact and diffuse substructures 
with masses on this scale have been derived from the analysis of data from stellar 
surveys of the Andromeda galaxy in light of gravitational-microlensing 
considerations~\cite{Niikura:2017zjd,Croon:2020ouk}.  Future pulsar-timing arrays 
(PTAs) could also in principle be sensitive to both compact and diffuse substructures on 
this mass scale --- and potentially on mass scales down to $M \sim 10^{-13}$ due to the 
characteristic imprints they would leave on the timing residuals obtained for a set of 
pulsars at the population level~\cite{Ramani:2020hdo,Lee:2021zqw}.  Substructures of this 
mass correspond to wavenumbers $k_{\rm phys} \sim 10^9~{\rm Mpc}^{-1}$.

While dark-matter substructures of this sort can arise as a consequence of either an epoch
of matter/radiation stasis or an EMDE, the results shown in Fig.~\ref{fig:Rsq} indicate that 
the morphology of the feature in $P(k,t_{\rm now})$ which results from such a stasis epoch can 
differ appreciably from that of the feature which results from an EMDE --- especially in 
situations in which $\barOmega_\gamma$ is non-negligible.  These differences in 
$P(k,t_{\rm now})$ can in principle lead to population-level differences in the properties 
of dark-matter substructures within galaxies and clusters.  Such differences could in principle 
be used in order to determine whether a population of dark-matter substructures might have
originated from a stasis epoch or from an EMDE.  We leave the investigation of this possibility 
for future work.

%%%%%%%%%%%%%%%%%%%%%%%%%%%%%%%%%%%%%%%%%%%%%%%%%%%%%%%%%%%%%%%%%%%%%%%%%%%%%%%%
%\FloatBarrier
\section{Conclusions\label{sec:conc}}

%%%%%%%%%%%%%%%%%%%%%%%%%%%%%%%%%%%%%%%%%%%%%%%%%%%%%%%%%%%%%%%%%%%%%%%%%%%%%%%%

In this paper, we have analyzed the behavior of density perturbations during 
a stasis epoch.  We have demonstrated that density perturbations for a spectator 
component --- \ie, a component which couples to the stasis sector only through 
gravity --- evolve differently during a stasis epoch than they would during an epoch in which
the universe is dominated by a single perfect fluid, even if the effective equation-of-state 
parameter $\langle w\rangle$ for both universes is the same.
In particular, we have shown that density perturbations associated with a spectator matter component exhibit 
power-law growth during stasis after horizon entry, with a power-law exponent which is determined 
solely by the value of $\langle w \rangle$.  As a result, a stasis epoch prior to BBN can 
lead to an enhancement in small-scale structure reminiscent of that 
which results from an EMDE, but with the magnitude of that enhancement depending on the balance
between the abundances of the cosmological components which participate in the
stasis.   This provides a sharp observational handle that can be used to distinguish a stasis epoch 
from more traditional cosmological epochs. The phenomenological consequences of this observation are 
currently under study~\cite{perturb_stasis_2}.

A few comments are in order.  First, for simplicity, we have taken the effective
sound speed for the radiation component to be $c_{s\gamma} \approx 1/\sqrt{3}$.  In scenarios 
wherein additional effects are present which modify $c_{s\gamma}$, the evolution of cosmological
perturbations becomes more complicated.  However, the qualitative results that we have
obtained are unchanged within the regime in which the impact that these effects have on 
$c_{s\gamma}$ is small.

Second, in this paper, we have assumed for simplicity 
that the perturbations for all of the fields in this scenario, including the spectator, are 
initially purely adiabatic.  Such initial conditions arise naturally in scenarios in which the
cosmological populations of $\chi$ and all of the $\phi_\ell$ are generated predominately either 
directly via inflaton decay during reheating or subsequently via interactions among particles 
in the radiation bath --- \eg, via freeze-in or freeze-out.  By contrast, such initial conditions 
are less natural in scenarios in which these populations of $\chi$ and/or the $\phi_\ell$ are 
generated predominately via processes such as vacuum-misalignment production, wherein the 
primordial fluctuations of these fields are uncorrelated with those of the inflaton.
Nevertheless, we emphasize that even in realizations of matter/radiation stasis wherein the 
initial conditions for $\chi$ and the $\phi_\ell$ are not adiabatic, the right side of
Eq.~(\ref{eq:dPrimePhiEqnArbwSimpFTDimVars}) is still in general non-vanishing, and thus 
$\delta_{k\chi}$ still evolves differently than it does during an epoch of perfect-fluid 
domination.

Many extensions of this work are also possible.  For example, in this paper we have 
studied the implications of a matter/radiation stasis, but since any realization or stasis generically 
gives rise to a non-vanishing source term for $\Phi_k$ in Eq.~(\ref{eq:dPrimePhiEqnArbwSimpFTDimVars}),  
we expect that similar results will emerge from other realizations of stasis as well --- including 
those realizations in which the stasis involves more than two cosmological energy components.  The effects 
we have uncovered here may also have an impact on the spectrum of primordial perturbations 
which emerges from models of stasis inflation~\cite{Dienes:2024wnu}.

%%%%%%%%%%%%%%%%%%%%%%%%%%%%%%%%%%%%%%%%%%%%%%%%%%%%%%%%%%%%%%%%%%%%%%%%%%%%%%%%

\section*{Acknowledgements}

The research activities of KRD are supported 
in part by the U.S.\ Department of Energy under Grant DE-FG02-13ER41976 / DE-SC0009913, and 
also by the U.S.\ National Science Foundation through its employee IR/D program.
The research activities of LH are supported by the STFC under Grant ST/X000753/1.  
The research activities of DH are supported in part by the U.S.\ National Science 
Foundation under Grant PHY-2014104.  The research activities of FH are supported in part 
by the Israel Science Foundation grant 1784/20, by MINERVA Grant 714123, and by 
MINERVA Grant, Project 7141230301.  
The research activities of AP and BT are supported in part by the U.S.\ National Science 
Foundation under Grants PHY-2014104 and PHY-2310622.  The opinions and conclusions
expressed herein are those of the authors, and do not represent any funding agencies. 

%\appendix 

%\clearpage

\bibliography{references}

\end{document}